# A short-range metastable defect in the double layer ice

Zhiyuan Zhang[1], Yu Zhu[1], Le Jin[1], Xinrui Yang[1,2], Yanchao Wang[2,3], Chang Q. Sun[4*] and Zhigang Wang[1*]

[1]*Institute of Atomic and Molecular Physics, Jilin University, Changchun 130012, China*

[2] *College of Physics, Jilin University, Changchun 130012, China*

[3] *International Center for Computational Methods and Software & State Key Lab of Superhard Materials, Changchun 130012, China.*

[4]*School of Electrical and Electronic Engineering, Nanyang TechnologicalUniversity,639798, Singapore.*

*E-mails: wangzg@jlu.edu.cn (Z. W.); ecqsun@ntu.edu.sg (C. S.)

**Abstract**

Although the phase of water has extensively investigated whether there exists a defect distorting only locally the structure still under debate. Here we report a localized 5775 defect phase presented in the double layer ice on the Au (111) surface, which is a metastable structure with 5- and 7-membered rings compared with a perfect hexagonal one. Without altering the total number of the hydrogen bonds of the ice, the defect only introduces 0.08 Å molecular displacement and 3.27% interaction energy change outside the defected area. Such defect also exists without Au support but causes a larger lattice relaxation or smaller interaction energy change. The excessively high barrier as well as the low quantum tunneling and thermodynamic probabilities hinder the formation of the defect by post-grown isomerization from the perfect to the defected structure. This finding indicates that the defected ice is stable, and the defect can be formed during the ice growth stage.

**Main text**

Much attention has been paid to the phases of water since the last century[1-3]. Apart from the regular phases of vapor, liquid and solid[4-6], some peculiar phases such as superionic[7], terapascal[8], negative-pressure[9] and quasisolid phase[10] have been discovered, particularly at the surface and interface. Water molecules can form either layered structures of long-order on graphene[3], mica[11], Pt[12], and Au[13,14] surfaces or with involvement of the Bjerrum defect[15,16] resulting in long-range disorder from the molecular rotation induced repulsive interactions[17]. However, whether there exist local defects that lead to the smallest disorder of long-range structures remains to be resolved, which is one of the basic attention-worthy questions in understanding the phase structure of water and ice.

We uncovered such an intrinsic defect consisting of two pairs of 5- and 7-membered ring in the double-layer ice that reserves the long-range order of ice on Au (111) surface or even in the isolated state from the first-principles calculations. The graphene-like 5775 defect with Stone-Wales typed of structure only causes short-range distortion without affecting long-range consistency. Revealing the properties of such defected phase can abundant the understanding of water system and provide a new point of view to the phase-related research.

**Results and Discussion**

Firstly, we obtain the two-dimensional double-layer ice by interlocking two layers of molecules on Au (111) surface. Structural optimization shows that O atoms outline the honeycomb-like hexagons in each layer, as shown in Fig. 1a. The thickness of each plane is about 0.10 Å thick. The average intra-layer O···O distance is 2.88 Å, and that of the interlayer is 2.77 Å. Every other water molecule in the planes have one O-H bond pointing to the other plane, thus forming 4-membered squares between these two layers of molecules in a clockwise or a counterclockwise manner. The shortest distance between Au atom and the O in the first layer of ice is around 3.39 Å apart.

A defect contains two pairs of adjacent 5- and 7-membered rings is discovered in the double layers (see in Fig. 1a), which is similar to the Stone-Wales type 5775 point defect in the carbon systems[18]. The defected area consists of two adjacent pairs of water molecules (the center), their neighboring 14 pairs at the its edge (the neighbor). The rest molecules outside the area remain as they are with negligible disorder (the outer). The defect is a metastable structure with an energy of 0.50 eV higher than that of the stable perfect one. In contrast to the Bjerrum defect with 120° rotation of single molecule in ice that derives the repulsive interactions, water molecules in such metastable structure reserves a four-coordinated H-bond network, which is insufficient to cause repulsive forces to destabilize the lattice. Calculations performed on the defected ice without the Au support reveal a similar four-coordinated structure with local dislocation, except for an average 0.19 Å lattice expansion in the 5-membered ring and 0.24 Å lattice contraction in the 7-membered ring. The energy of the defected structure is 0.41 eV higher than that of the perfect one.

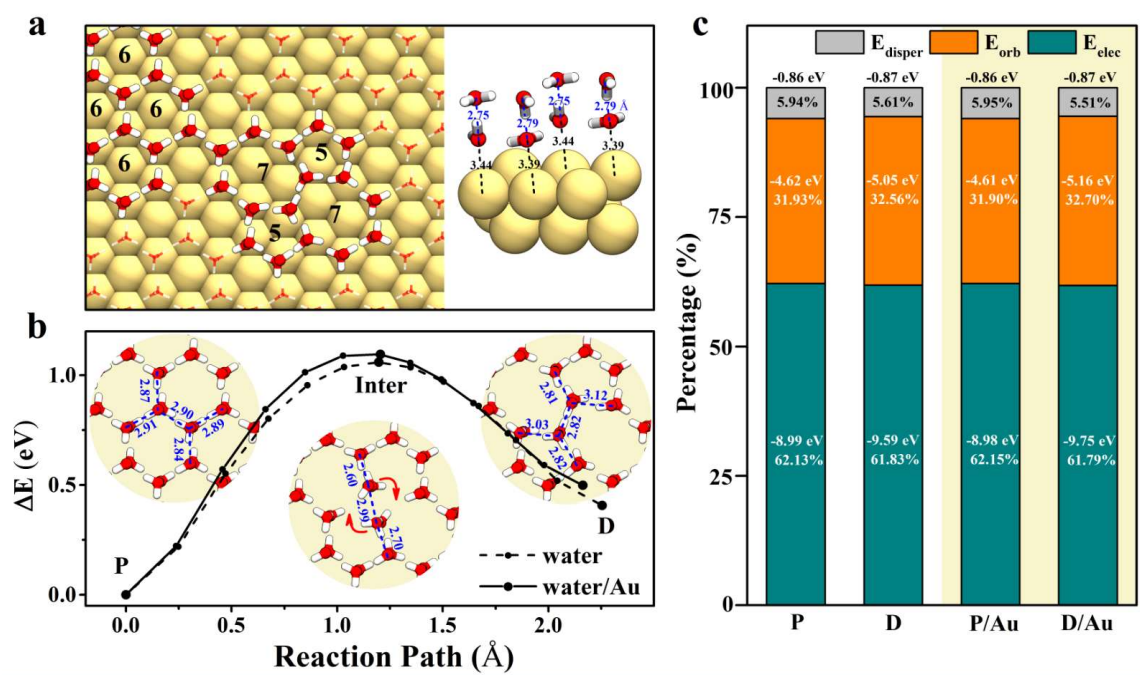

**Fig. 1 Structural and energetics of the double layer ice with and without Au (111) support. a**, Top view of the double layer ice on Au with the 5775 defect (left) and the detail of the interlocked structure (right). The red sphere and the white tube are O and H atom, deposited on top of the golden Au atoms. The highlighted water molecules are the defected area of 5- and 7-membered rings right in the middle and part of the perfect hexagons in the upper left. **b,** Potential energy curve along the rotation isomerization path connecting the perfect and the defect. The vertical axis is the relative energy to that of the perfect structure, and the horizontal axis is the equivalent coordinate of the rotation path. The insets are close views of the perfect (P), the intermediate (Inter) and the defect (D) structures. The blue and black numbers in **a** and **b** are O···O and O···Au distances (Å). **c**, The percentages for electrostatic ($E_{elec}$), orbital ($E_{orb}$) and dispersion ($E_{disper}$) in sum of these attractive interaction energy $E_{elec}$ + $E_{orb}$ + $E_{disper}$, while the repulsion energy ($E_{repul}$) is not presented in the figure.

Compared with the perfect structure, the center molecules displace by 1.90 Å in the defected area. The neighbor dislocates by 0.25 Å and those outside the area by 0.08 Å. For the unsupported structure, the displacements of the center and the neighbor are 1.96 Å and 0.41 Å, and that of the outer is 0.18 Å, which are all relatively larger than the case on the Au surface. The displacement for the outer of the defected structure on Au takes about 2.78% of the average O···O distance much lower than the 8.71% of the neighbor. For the unsupported system, the lattice distortions of the outer and the neighbor are 6.25% and 14.29%, respectively.

The energy decomposition analysis (EDA) is carried out between the defected area and the outer to examine the influence of the 5775 defect on the interacting feature. The result suggests that, after the formation of the defect in the Au-supported (unsupported) structure, both electronic ($E_{elec}$) and orbital ($E_{orb}$) attraction energy increase from -8.98 (-8.99) and -4.62 (-4.62) eV to -9.75 (-9.59) and -5.16 (-5.05) eV. Meanwhile, the repulsion energy ($E_{repul}$) increases from 6.50 (6.51) to 7.57 (7.35) eV, resulting in 0.20 (0.26) eV increase of total interaction energy ($E_{total}$) after the formation of the defect for ice with (without) Au support, which takes 3.27% (2.51%) of $E_{total}$ (see in Fig. 1c). Although the interaction energies are different between systems with and without the defect, the percentage of the terms are very close. Among total attracting parts, $E_{elect}$ takes about 61% ~ 63%, and $E_{orb}$ accounts for 31% ~ 33%, while the $E_{disper}$ is 5% ~ 6%. Therefore, the defect does not change the interacting feature between the defected area and the outer.

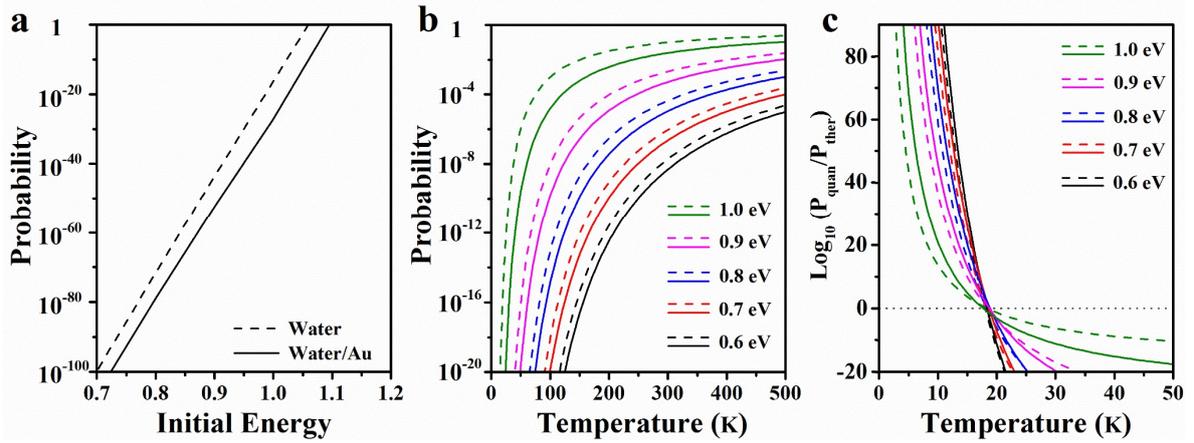

**Fig. 2 The probabilities of quantum tunneling, thermodynamics and their ratio for double layer ice with and without Au (111) support. a**, Quantum tunneling and **b**, thermodynamic probability with different energy and temperature. **c**, the ratio of quantum tunneling and thermodynamic probability under the same provided energies ($E_p$). The solid and dash lines in **a**, **b** and **c** represent the data of the defected double layer ice with and without Au support, respectively. The vertical axes are all logarithmic. The quantum tunneling probabilities in are obtained with Wenzel-Kramers-Brillouin (WKB) approximation, $P_{quan} = \mathrm{Exp}\left[-\frac{2}{\hbar}\int_{x_1}^{x_2}\sqrt{2m(V(x)-E_p)}\,dx\right]$, which is only relevant to the provided energy ($E_p$) for a certain potential surface V(x), where x is the coordinate in Fig. 1b. The $x_1$ and $x_2$ is the two intersects of $E = E_p$ and $E = V(x)$, respectively. The thermodynamic probability suits the Boltzmann distribution, $P_{ther} = \mathrm{Exp}\left(-\frac{\Delta E}{kT}\right)$, varying with $E_p$ and temperature T, where ∆E is the activation energy from the provided energy ($E_p$) to the barrier ($E_b$), $\Delta E = E_b - E_p$. The Ratio is obtained by $\mathrm{Ratio} = \left(\frac{P_{quan}}{P_{ther}}\right)$, and the horizontal dot line in **c** suggests that $P_{quan}$ and $P_{ther}$ are equal.

A rotation isomerization reaction connecting the perfect and the defected structures can be obtained through path searching[19]. It is found that the energy barrier of the system on the surface is 1.10 eV, while that without Au support is 1.06 eV. The barriers are much larger than the generally admitted bond energy of H-bond about 0.20 eV[20]. To overcome the barrier, the rotation path can be achieved through quantum tunneling or thermodynamic approach. In these approaches, a provided energy ($E_p$) should be given higher than the energies of the perfect and defected structure, thus giving out probabilities of quantum tunneling ($P_{quan}$) and thermodynamics ($P_{ther}$). The result shows that $P_{quan}$ is extremely low, which is lower than $10^{-10}$ even with an $E_p$ of 1.00 eV. In comparison, when temperature is higher than about 20 K (see Fig. 2c), thermodynamic transport is preferred. However, the temperature must be over 300 K with $E_p$ of 1.00 eV to reach a non-negligible $P_{ther}$ higher than 10% to overcome the barrier, while under the experimentally observed existing temperature at about 120 K[13], the $P_{ther}$ is lower than $10^{-4}$.

**Conclusion**

In this study, we have found, for the first time, that a short-range 5775 typed defect can exist in the double layer ice on Au (111) surface. Removing the Au substrate will not essentially change the structural configurations, indicating that the existence of the defect does not rely on the substrate. The introduction of the defect will neither alter the overall four-coordinated H-bond network nor change the interacting feature between the molecules, maintaining the stability of the system. The defect only leads to negligible difference in the interaction energy and atomic dislocation outside the defected area, showing that the defect only causes local distortion. The small difference of the interaction energy component suggest that the defect does not change the interacting feature. These results suggest that the 5775 defect is short-range defect that reserves the long-range consistency of the double layer ice. The high energy barrier for the rotation isomerization path connecting the perfect and the defected structure is nearly impossible to overcome for the system under its existing temperature. The the low quantum tunneling and thermodynamic probabilities also hinders the possibility of such reaction. Experimental research has found that 5- and 7-membered rings could be observed at the edge of ice sheet, and gradually grow into hexagonal rings as the sheet grows[13]. Instead of the rotation reaction, the defect could be formed from the non-ideal distortion of this process.

**Methods**

All structural and energetic calculations are carried out using density functional theory (DFT). The structural optimization and the rotation path are obtained with Perdew-Burke-Ernzerhof (PBE)[21] level, which is generally used for describing water related systems. The cut off energy are set to 340 eV. The above calculations are adopted with CASTEP package[22]. The double layer containing 64 water molecules on two layers of Au (111) surface is modeled in a large 17.30 Å×19.98 Å×15.00 Å periodic cuboid cell. Energy decomposition analysis (EDA) are performed based on Morokuma's method[23-25] to examine the influence of the defect on the interacting feature. Two approaches to cross the barrier of the rotation path, the quantum tunneling following Wenzel-Kramers-Brillouin (WKB) approximation[26-29] and the thermodynamics suiting Boltzmann distribution, are analyzed to explore the possibility of the rotation path connecting a perfect structure to the defected one are examined.

**Acknowledgement**

The authors would like to acknowledge Dr. Depeng Zhang, Rui Wang, Danhui Li and Aihua Cheng for constructive discussions. Z. W. would also acknowledge the National Natural Science Foundation of China (Nos.: 11674123 and 11974136), and the High-Performance Computing Center of Jilin University and National Supercomputing Center in Shanghai.